\documentclass[pdflatex,sn-mathphys-num]{sn-jnl}

\usepackage{graphicx}%
\usepackage{multirow}%
\usepackage{amsmath,amssymb,amsfonts}%
\usepackage{amsthm}%
\usepackage[version=4]{mhchem}
\usepackage{mathrsfs}%
\usepackage[title]{appendix}%
\usepackage{xcolor}%
\usepackage{textcomp}%
\usepackage{manyfoot}%
\usepackage{booktabs}%
\usepackage{algorithm}%
\usepackage{algorithmicx}%
\usepackage{algpseudocode}%
\usepackage{listings}%
\usepackage{indentfirst}
\usepackage{comment}

\graphicspath{{figures/}}

\begin{document}

\title[Article Title]{TBHubbard: tight-binding and extended Hubbard model database for metal-organic frameworks}


\author[1]{\fnm{Pamela} \sur{C. Carvalho}}
\equalcont{These authors contributed equally to this work.}

\author[2,3]{\fnm{Federico} \sur{Zipoli}}
\equalcont{These authors contributed equally to this work.}

\author[4]{\fnm{Alan} \sur{C. Duriez}}

\author[4]{\fnm{Marco Antonio} \sur{Barroca}}

\author[4]{\fnm{Rodrigo} \sur{Neumann Barros Ferreira}}

\author[5]{\fnm{Barbara} \sur{Jones}}

\author[6]{\fnm{Benjamin} \sur{Wunsch}}

\author*[4]{\fnm{Mathias} \sur{Steiner}}\email{mathiast@br.ibm.com}

\affil[1]{\orgname{IBM Research}, \orgaddress{\city{São Paulo}, \postcode{04007-900}, \state{SP}, \country{Brazil}}}

\affil[2]{\orgname{IBM Research Europe}, \orgaddress{\city{Saümerstrasse 4}, \postcode{8803 Rüschlikon}, \state{Zurich}, \country{Switzerland}}}

\affil[3]{\orgname{National Center for Competence in Research-Catalysis (NCCR-Catalysis)},  \state{Zurich}, \country{Switzerland}}

\affil[4]{\orgname{IBM Research}, \orgaddress{\city{Rio de Janeiro}, \postcode{20031-170}, \state{RJ}, \country{Brazil}}}

\affil[5]{\orgname{IBM Quantum}, \orgname{IBM Research Almaden}, \orgaddress{\city{San Jose}, \postcode{95120}, \state{CA}, \country{USA}}}

\affil[6]{\orgname{IBM Research}, \orgname{IBM T.J. Watson Research Center, Yorktown Heights}, \orgaddress{\city{New York}, \postcode{10598}, \state{NY}, \country{USA}}}



\abstract{

Metal-organic frameworks (MOFs) are porous materials composed of metal ions and organic linkers. Due to their chemical diversity, MOFs can support a broad range of applications in chemical separations. However, the vast amount of structural compositions encoded in crystallographic information files complicates application-oriented, computational screening and design. The existing crystallographic data, therefore, requires augmentation by simulated data so that suitable descriptors for machine-learning and quantum computing tasks become available. Here, we provide extensive simulation data augmentation for MOFs within the QMOF database. We have applied a tight-binding, lattice Hamiltonian and density functional theory to MOFs for performing electronic structure calculations. Specifically, we provide a tight-binding representation of 10,000 MOFs, and an Extended Hubbard model representation for a sub-set of 240 MOFs containing transition metals, where intra-site $U$ and inter-site $V$  parameters are computed self-consistently. The data supports computational workflows for identifying structure-property correlations that are needed for inverse material design. For validation and reuse, we have made the data available at \href{https://dataverse.harvard.edu/dataverse/tbhubbard/}{https://dataverse.harvard.edu/dataverse/tbhubbard/}.
}

\maketitle

\section*{Background \& Summary}

Metal-organic frameworks (MOFs) are porous materials with applications in gas capture and storage \cite{HaoLi2018, Zhao2024}, catalysis \cite{Iliescu2022, Bavykina2020}, biomedicine \cite{Sezgin2025, AbanadesLazaro2024},  electrical conductivity \cite{Check2025, Xie2020}, transport and diffusion \cite{Fujie2015, WuZhendi2023} and chemical sensing \cite{Kreno2012,KuanChang2023}. They consist of structural building blocks formed by metal clusters and organic linkers, and form ordered, nanoscale pores \cite{Yusuf2022}. Their properties are determined by the combination of their building units under a certain topology, which is unique to each structure.

Given the wide variety of MOFs that have been hypothesized~\cite{lee2021computational} and synthesized~\cite{moghadam2017development}, high-throughput computational screening plays an important role in identifying MOF candidates that are suitable for a specific application. Even though screening MOF databases by means of ab-initio simulations is feasible, see \cite{Mancuso2020}, it is computationally unpractical in many cases. Data-driven techniques can aid in processing large amounts of MOF data by identifying correlations between structure and property in machine-learning (ML) based workflows \cite{YutongLiu2025}. Likewise, computational discovery by means of inverse design with pre-defined  target properties  \cite{Han2025}, such as, for example, the creation of MOFs with high CO$_{2}$ affinity \cite{Park2024, Boyd2019, Jablonka2020}, requires the application of advanced ML methods and high-quality datasets. 

To effectively screen and design MOFs for applications, both structural and electronic data are needed. While structural data files are publicly available in large quantities \cite{Zhao2025, Gibaldi2025, moghadam2017development}, electronic structure data are sparse. For MOFs, high-quality electronic structure data could be created by deploying a suitable physical model that accounts for electronic correlations in the presence of metal clusters. 

Computationally, PAOFLOW \cite{Nardelli2018, Cerasoli2021} is a key tool for projecting the electronic structure into a tight-binding (TB) Hamiltonian using localized valence atomic orbitals as a basis. Thus, nearest-neighbors interactions are captured by the matrix coefficients which could potentially be used in ML workflows to predict new structures with optimized features.

\begin{figure}[H] 
    \centering
    \includegraphics[width=\textwidth]{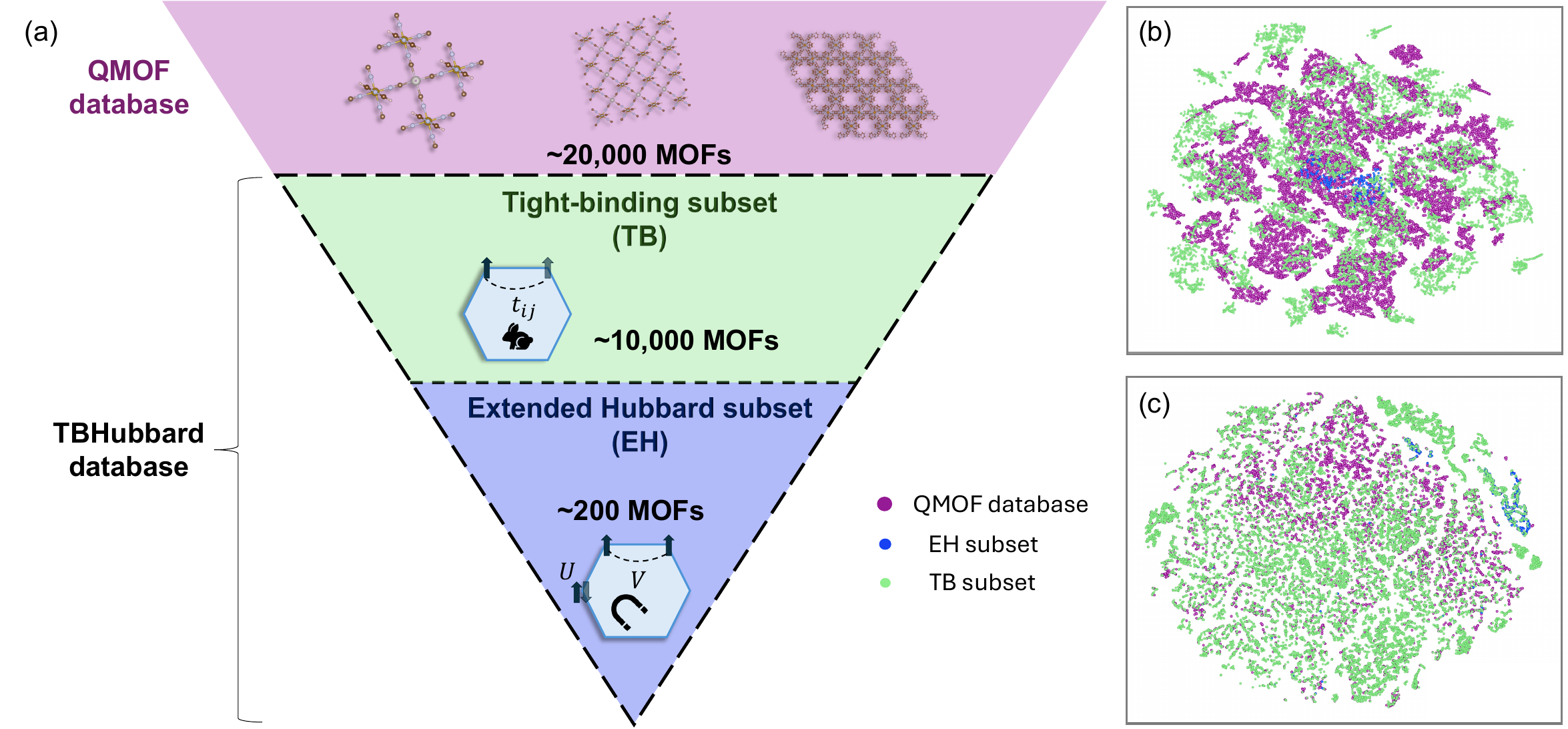} 
    \caption{(a) Illustration of the TBHubbard dataset. The QMOF~\cite{Rosen2021} database is indicated in pink, providing over 20,000 MOF structures. From this data collection, the TBHubbard database comprises two subsets of materials: the Tight-binding (in green) and Extended Hubbard (in blue) subsets with $\approx$ 10,000 and  $\approx$ 200 materials, respectively;
    (b) t-SNE projection of tight-binding matrices, where points are colored according to the different databases analyzed in this study;
    (c) t-SNE projection of SOAP-3~\AA{} descriptors for metal atoms across the dataset. A preliminary PCA step reduced the descriptor dimensionality to 8 components, retaining 97~\% of the total variance. 
    The color scheme for the t-SNE plots is as follows: pink for the QMOF database, blue for the EH subset, and green for the TB subset.
    }
    \label{fig:intro}
\end{figure}

For the purpose of this contribution, we consider a lattice Hamiltonian with correctional terms to properly account for the electronic correlations and hybridization occurring within and between the building units of representative MOFs. In the Extended Hubbard (EH) model \cite{Campo2010}, strong electronic correlations are captured by the intra-site and inter-site Hubbard parameters $U$ and $V$, respectively. Typically, $U$ corrects for transition metal contributions, due to their localized atomic orbitals, while $V$ represents the interaction between metal ions and their nearest neighbors. In the case of MOFs, we expect these parameters to improve upon standard tight-binding (TB) description by accounting for the electronic interaction within metal clusters and the hybridization with organic constituents.

An important application of the EH model in electronic structure calculations based on density-functional theory (DFT) is introducing corrections to \textit{d} and \textit{f}-electron energies. Known as DFT+U or DFT+U+V, these methods improve the accuracy of band gap predictions where standard DFT typically fails \cite{Pavarini2021}. The $U$ parameter is often estimated empirically, however, it can be derived from first-principles \cite{Mann2016, Timrov2022}. As such predictions are computationally costly, there is a growing need for creating datasets of Hubbard parameters \cite{Yu2020,Uhrin2025}.
In the case of MOFs, a dataset augmentation with computed $U$ and $V$ values would be a valuable  contribution to the theoretical investigation of MOFs. In particular, it would enable the application of data-driven strategies in materials screening and design workflows.

In this work, we provide two augmented datasets based on QMOF \cite{Andrew2021, Rosen2022} that are represented visualized in Fig.~\ref{fig:intro}. In the TB dataset, we have projected the electronic density onto a tight-binding Hamiltonian with PAOFLOW \cite{Nardelli2018, Cerasoli2021}, and provided TB matrices as well as Smooth Overlap of Atomic Positions (SOAP) \cite{Bartok2013} descriptors for 10,435 materials. In the EH dataset, besides the TB parameters, we computed intra-site $U$ and inter-site $V$ Hubbard parameters for a select group of 242 MOFs using Quantum Espresso \cite{Gianozzi2009,Gianozzi2017,Gianozzi2020}, enabling the creation of an Extended Hubbard Hamiltonian representing each material. The datasets allow for exploring potential correlations between structural and electronic properties in computational workflows for MOF screening and design.

\section*{Methods}

We provide two complementary subsets tailored for inverse design: 
TB and EH subsets. In the case of the TB subset, the first-principles DFT calculations are performed for 10,435 MOFs along with their TB projection and Smooth Overlap of Atomic Positions (SOAP)  descriptors, which are used as fingerprints of the local environment to describe the topology of each material. While the DFT-based calculations provide a quantum-mechanical foundation for electronic structure analysis, the SOAP descriptors offers a data-driven representation of atomic environments, enabling material discovery. In the case of the EH, besides performing the TB projection, additional EH parameters, i.e. intra-site $U$ and inter-site $V$ interactions, have also been computed for a smaller set of 242 MOFs. In the following, we outline the methodologies used for generating the data. This includes selecting the structures, performing the ground-state calculations and tight-binding projections, as well as computing the SOAP descriptors, TB embeddings and Hubbard parameters.

\par \textbf{Structure Selection.} Our starting point is the QMOF database~\cite{Andrew2021, Rosen2022}, containing 20,375 metal-organic framework structures. The selection process for creating the TB subset begins by down-sampling from 20,375 to 10,435 MOFs, prioritizing the diversity of metal ions in the clusters and focusing on structures without spin polarization. 
In a next step, we have filtered the data set down to 242 MOFs to form the EH subset. The step involves selecting materials with a pore-limiting diameter (PLD) larger than 3.3 \AA~ for maximizing the application potential, including \ce{CO2} capture and ionic transport. 
For the subset, we have focused on materials with at least one transition metal in the cluster. Note, that we have avoided the computation of $U$ for metals such as Zn, Cd, Hg, Cn and La which has led to produce nonphysical values for $U > 20~\text{eV}$.

\par \textbf{Ground State.} To perform electronic structure calculations based on DFT, we have used the Quantum Espresso (QE) software \cite{Gianozzi2009,Gianozzi2017,Gianozzi2020}, within the generalized gradient approximation (GGA) \cite{Perdew1996}.
The inputs of our calculations, such as atomic positions, spin-polarization and \textbf{k}-point mesh, resemble those available in the QMOF database.
Following previous work related to \textit{ab initio} calculations of MOFs \cite{Mancuso2020}, we have included Grime's D3 Van der Waals corrections \cite{Grimme2010} with zero damping (\texttt{dftd3 = 4}). The values used as convergence threshold and mixing factor for self-consistency are $10^{-8}$ and 0.1, respectively, along with Davidson diagonalization. We have fixed the atomic positions throughout the self-consistent cycle, considering that the structure's geometries were previously optimized. For the composition of each material, we have used the hardest pseudopotential among the species involved as kinetic energy cutoff, except when unavailable. In that case, we have used 50 Ry and 400 Ry for wavefunction and density, respectively. 

For the TB subset, to ensure the $\Gamma$ point is included for all materials, we have ignored shifted \textbf{k}-point meshes. Nevertheless, we have validated for a few, representative materials that the use of shifted/unshifted meshes did not alter the total energy by more than $\sim$ 10 $\mu$eV. The van der Waals (vdW) correction has been excluded from the SCF calculations, since they do not affect the wavefunction analysis and projection of the density of states. As a result, the total energy does not include any vdW contributions.
Given the broad range of elements involved, we have chosen ultrasoft pseudopotentials from the PSlibrary \cite{DalCorso2014} to ensure compatibility with a diverse set of materials. The electronic density for computing Bader charges has been exported in cube format, as they are commonly used as descriptors of atomic structure and features.

For the EH subset, we have used pseudopotentials taken from the SSSP PBE Efficiency v1.3.0 set \cite{Prandini2018}, taking into account that the prediction of Hubbard parameters are computationally expensive. Note, that the ground-state computations have been carried out considering both initial, negligible guesses of Hubbard parameters, of the order of $10^{-8}~\text{eV}$, which are necessary for the application of density-functional perturbation theory, in addition to self-consistently computed $U$- and $V$-values.
At this point, we have computed the band gaps as the energy difference between the lowest and highest occupied levels. The band-gap values computed with DFT+U+V for \textit{d-p} and \textit{d-s} perturbations differ, as the computations are performed with different sets of $U$- and $V$-values.

\par
\begin{figure}[H] 
    \centering
    \includegraphics[width=1.\textwidth]{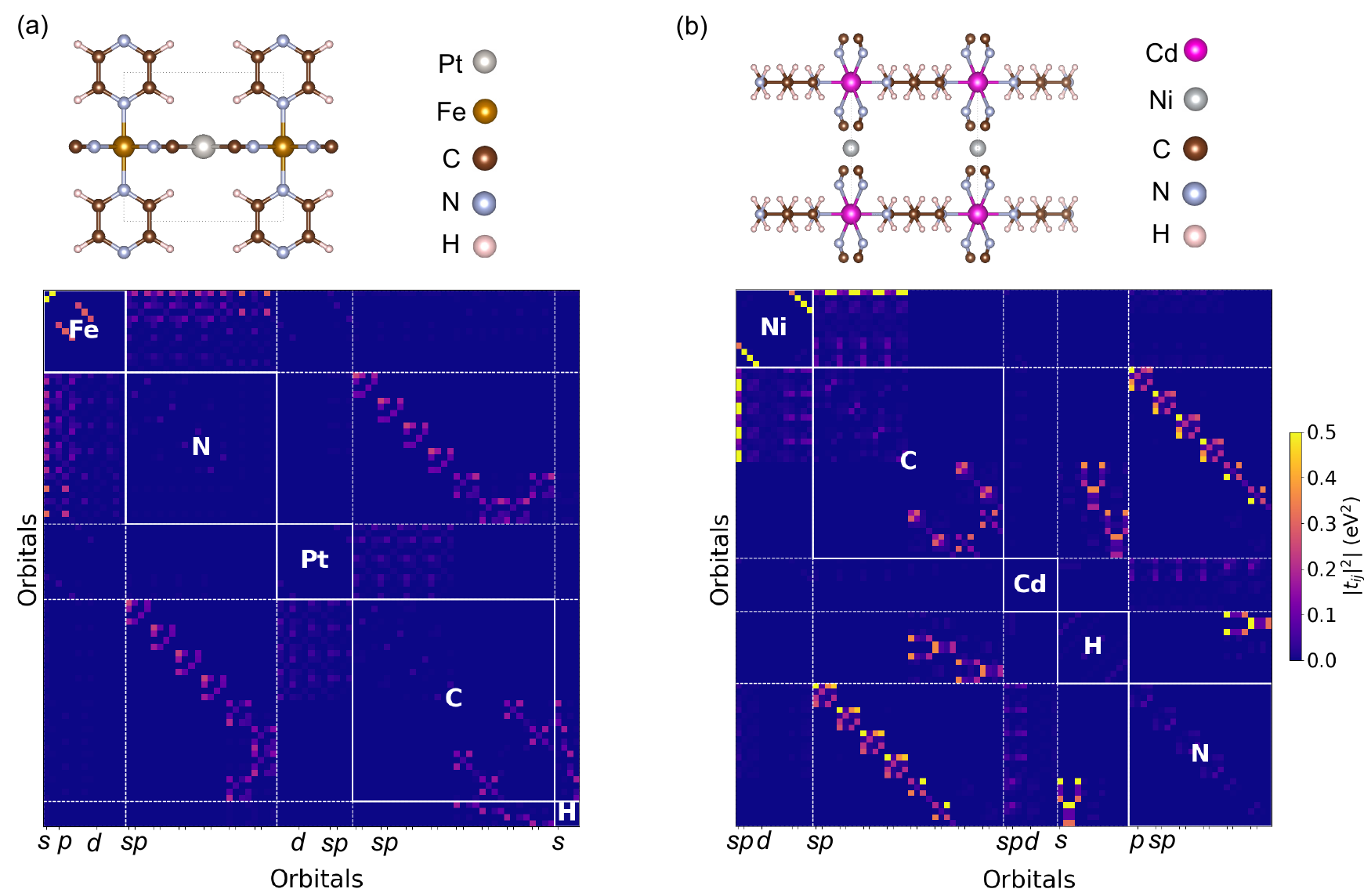} 
    \caption{False-color image representing the normalized $|t_{ij}|^2$ tight-binding matrix coefficients of the localized orbital basis set for MOFs (a) \ce{FePtC8H4N6} (or \texttt{qmof-3dfbcbd}) and (b) \ce{CdNiC8H12N6} (or \texttt{qmof-4d9a98c}), with their respective structural representations shown above. For visualizing the matrix, the maximum intensity is set to 0.5 and the matrix diagonal to 0. The MOF structure images were created using VESTA \cite{Momma2008}.}
    \label{fig:hopping}
\end{figure}

 \textbf{Tight-binding Projection.} 
In order to obtain a tight-binding representation of the selected materials, we have used PAOFLOW \cite{Nardelli2018, Cerasoli2021}, a software tool which embeds outputs from electronic structure, plane-wave pseudopotential calculations into pseudo-atomic orbitals. The projection provides the Hamiltonian, both in real and reciprocal space, within the localized orthogonal atomic basis according to the valence orbitals present in the pseudopotential used for each element. The Hamiltonian coefficients are known as TB parameters. PAOFLOW is fully compatible with QE, which facilitates the data conversion, as only the output of the self-consistent part is required for the projection. In the present subset, we have computed the tight-binding coefficients for each material based on the standard DFT output and the information of the corresponding atomic orbital basis. In Fig.~\ref{fig:hopping}, we show examples of the TB matrix visualization for two representative materials. 

\par \textbf{Machine-learning Descriptors.}
To characterize the structure of MOFs, we have computed two complementary descriptor sets. The first is based on SOAP descriptors, which encode local atomic environments. The second is based on an original contribution of this work: a DFT-informed fingerprint derived from TB projections and structured into fixed-size embeddings that capture essential features of the electronic structure.

Analysis of the local environment of the metal clusters in MOFs is crucial, and we have anonymized the identity  of metal atoms in our SOAP descriptor calculations. 
We have adopted a reduced species set that merges metals within a single category, while maintaining detailed classifications for the key elements of most organic linkers, such as hydrogen, oxygen, carbon, and nitrogen. This simplification is especially useful when dealing with large datasets. 
For the purpose of this contribution, we have used two sets of SOAP descriptors. The SOAP-3~\AA~ descriptor captures information about the first nearest-neighbor shell while the SOAP-5~\AA~ descriptor extends beyond the second shell, thus providing a broader range of local structural information. While the former contains 684 values, the latter has 1500 values.

For creating the TB embeddings, we have standardized the TB matrix entries for each atom pair and cast them into a fixed-size format to enable consistent application in machine-learning models. Each $t_{ij}$ between atomic orbitals is embedded into a $13 \times 13$ matrix block, regardless of the specific elements involved. The fixed shape is designed to accommodate the maximum number of relevant orbital combinations commonly found in MOFs.

The $13 \times 13$ matrix stems from setting up the orbital configuration per element as 2 $s$ orbitals, e.g., semicore and valence, 6 $p$ orbitals, i.e., 3 semicore and 3 valence, as well as 5 $d$ orbitals. This leads to $13 \times 13$ blocks that captures all possible, pairwise orbital interactions between two atoms. The matrix is filled with zeros in case orbitals are absent from a specific element.

We have excluded $f$-orbitals from this basis due to their relative scarcity in the systems studied and due to the significant increase in matrix dimensionality they would entail, up to $20 \times 20$. This would more than double the computational burden and potentially dilute the machine-learning correlation signal of the prevalent orbital types.
In cases where a pseudopotential provides fewer orbitals, e.g., only a single $s$ or $p$ orbital, we map the available orbitals to the \emph{outermost valence orbitals} within the standardized block structure. This ensures that the most chemically relevant contributions, i.e., those related to bonding and reactivity, are retained and aligned consistently across different elements. In spin-polarized systems, the matrix size could be doubled to account for spin-up and spin-down channels. While not in scope of the present work, this would not affect the generalization of the approach, unlike the inclusion of $f$-orbitals which would significantly increase the representation size.
For each atom pair $(i, j)$, a slice of the full TB matrix is extracted, corresponding to their orbital interaction block. The $13 \times 13$ matrices are extracted from both diagonal and off-diagonal entries, ranked by the maximum absolute value of the strength of interaction, and the top-$k$ blocks are then selected. The resulting $13 \times 13 \times k$ tensor is flattened to form a fixed-length vector that represents the TB embedding for the atom. These embeddings are aligned with corresponding SOAP descriptors to serve as structured inputs for machine learning workflows.

In this context, the TB embeddings provide the source data, while the SOAP descriptors act as the target data. 
The embeddings provide a structured representation of a material's electronic environment and can be used to predict atomic pair species as well as SOAP descriptors, facilitating material characterization and structure reconstruction.

\par \textbf{Hubbard Parameters.} To compute the intra-site $U$ and inter-site V Hubbard parameters,  we have used the QE software based on the implementation of density functional perturbation theory in the DFT+U+V framework \cite{Campo2010,Timrov2022}. The $U$ correction applies to transition metals and $V$ represents the coupling between the transition metal and its nearest-neighbors. The Hubbard parameter calculation is orbital-dependent and the desired manifold should be defined \textit{a priori} in the self-consistent step. We have chosen the \textit{d} and \textit{d-p} orbitals as manifolds for $U$ and $V$, respectively, where the localized \textit{d} orbital is located at the transition metal and the \textit{p} orbital is located at the metal's nearest neighbor. Note, that it is not possible to compute Hubbard parameters for more than one manifold at a time, due to the high computational cost. We have selected a subset of 186 materials for computing $U$ and $V$ for \textit{d} and \textit{d-s} orbitals as manifolds. For identifying the different results for $U$ and $V$, we have adopted \textit{d-p} and \textit{d-s} perturbations as nomenclature following the chosen manifolds. For constructing the Hubbard projector, we have used Lowdin orthogonalized atomic orbitals and the parameters of \textbf{q}-grid were set to half of the \textbf{k}-point mesh used in the self-consistent step. For performing the perturbation, we have identified nonequivalent atoms by symmetry. Atoms of the same type, but not equivalent, are differentiated (\texttt{find\_atpert = 3})
and the convergence threshold for the response function $\chi$ was set to $10^{-7}$.
\newline

\section*{Data records}

The TBHubbard database is available in the Harvard Dataverse \href{https://dataverse.harvard.edu/dataverse/tbhubbard}{https://dataverse.harvard.edu/dataverse/tbhubbard} \cite{tbhubbard}. The calculations were performed in a high-performance computing cluster equipped with \texttt{x86} and \texttt{PowerPC} compute nodes. Computation time depended on simulation type, and the calculation of Hubbard parameters stands out as the most computationally intensive. Specifically, the 10,435 ground-state calculations took about 12 CPU-years while the 412 Hubbard parameter calculations took roughly 10 times that, about 127 CPU-years.

In total, the TBHubbard repository provides 10,435 ground state calculations, including the QE self-consistent input and output files. The PAOFLOW projection output is provided along with a serialized object file (\texttt{pickle}) containing additional information. We have included the list of $n_{orb}$ orbitals used as basis and the tight-binding Hamiltonian, in real and reciprocal space. The Hamiltonian is stored as tensor of dimensions [$n_{\text{orb}}, n_{\text{orb}}, k_1, k_2, k_3, 1$], where the \textbf{k}-point grid
($k_1 \times k_2 \times k_3$) is the Monkhorst-Pack grid used in the QE input. In the TB subset, the charge density computed with QE is also provided along with the Bader charge \cite{Henkelman2006}. The SOAP descriptors, TB embeddings and the \texttt{*.xyz} file with the atomic positions are also available. 
In addition, for the EH subset, we have provided 428 outputs from the \texttt{hp.x} QE executable containing the intra-site $U$ and inter-site $V$ Hubbard parameters, for which 186 refer to the \textit{d-s} perturbations and the remaining 242 are for \textit{d-p} perturbations. The results from the EH subset have been parsed and saved in a JSON format file, storing information such as structure, band gap and Hubbard parameter values.

\section*{Technical validation}

\begin{figure}[H] 
    \centering
    \includegraphics[width=1.\textwidth]{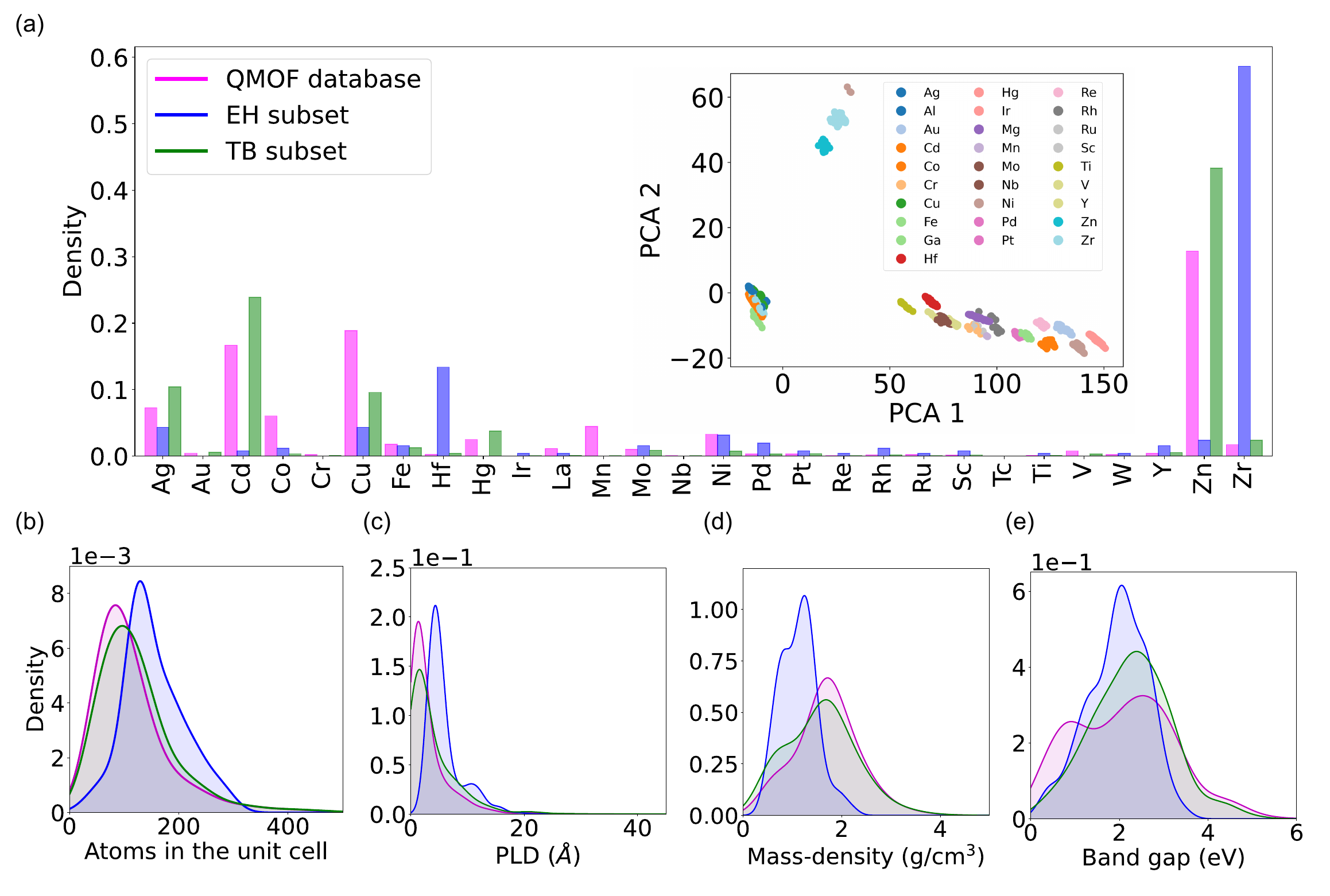}
    \caption{
    (a) Density histogram comparing the transition metal distribution in the QMOF database with the Tight-binding (TB) and Extended Hubbard (EH) data sets. The inset shows the PCA projection of TB embeddings for symmetry-independent metal atoms. In (b)-(e), probability density functions, comparing the QMOF database with the TB and EH data sets, are shown with regards to number of atoms in the unit cell; pore-limiting diameter, PLD, (in \AA); mass density (in g/cm$^3$) and standard DFT band gap (in eV), respectively.
    }
    \label{fig:stats_metals}
\end{figure}

\textbf{Diversity of structures.}
Before using first-principles simulation data for training machine learning models, let us investigate the chemical diversity of the data sets. We have created two distinct selections of materials data, each aiming at a different outcome. While the TB dataset is approximately half the size of the original QMOF database, the EH dataset contains, due to the higher computational cost, a much smaller number of materials. In Fig.~\ref{fig:stats_metals}(b)-(e), we present density plots of the materials distributions with regards to representative structural and electronic properties, for the QMOF database as well as the two data sets provided in this work. For all properties analyzed, the TB data exhibits a diversity distribution similar to the original database while the EH data differs, mainly due to the much smaller number of materials included. 

Another important feature is the distribution of transition metals, since their presence in metal-organic frameworks is essential to the application of the EH model. The density histogram for the two data sets and the QMOF database is plotted in  Fig.~\ref{fig:stats_metals}(a) for the structures that contain transition metal atoms. By design, the EH data set exhibits a concentration of MOF structures containing Zr and Hf atoms. Note, that Zr-based MOFs are of particular interest from and application perspective \cite{Ahmad2020, Sayed2022, Bai2016, Daliran2024, Gomez-Gualdron2014} since the discovery of UiO-66's high hydrothermal stability \cite{Cavka2008}. Similar structures have shown application potential in electrochemical sensors and biosensors \cite{Khosropour2024} as well as catalysts \cite{ZhangQiuyun2022, Abouseada2024}. While the inter-site $V$ parameters can provide insight into the metal-organic hybridization, the tight-binding parameters can be explored for investigating topological properties.

In the inset of Fig.~\ref{fig:stats_metals}(a), we present a two-dimensional Principal Component Analysis (PCA) \cite{Jolliffe2002} projection of the tight-binding (TB) embeddings. The explained variance ratios for the first two and first four principal components are 0.90 and 0.93, respectively, indicating that a large fraction of the variance is captured in low-dimensional space. Interestingly, clusters corresponding to different TB embeddings of the same metal emerge, suggesting that the embeddings capture metal-specific electronic structure characteristics.
In addition, for providing a broader view of the chemical diversity, we have included a t-distributed Stochastic Neighbor Embedding (t-SNE) plot \cite{vanDerMaaten2008}, constructed using the following features: number of atoms per unit cell, pore limiting diameter (PLD), largest cavity diameter, mass density, volume, band gap, and atomic number of the transition metal.
In Fig.~\ref{fig:intro}(b), we see a concentration of the EH data set, while the TB data set is spread out across the QMOF database, indicating a good representation of the source dataset. 
To visualize the diversity in the vicinity of metal atoms across the dataset, we computed a t-SNE projection of SOAP-3\AA~ descriptors, using PCA for the initial dimensionality reduction, see Figure~\ref{fig:intro}(c). The QMOF database as well as the EH and TB data sets form distinct clusters in the reduced space, suggesting that differences in structural motifs and generation protocols are well captured by the SOAP representation.

The diversity analysis is useful to differentiate the two data sets with regards to their application potential. While the TB projection provides topological data for exploring a broader range of metal-organic frameworks, the EH data provides Hubbard parameters for the focused investigation of Zr- and Hf-based MOFs. 
\newline

\textbf{Tight-binding matrix.}
By using the PAOFLOW software, we are able to represent metal-organic framework structures with tight-binding Hamiltonians and project the electronic density onto localized atomic orbitals. We have performed the projection for each \textbf{k}-point in the grid used in the ground state calculation, where the Hamiltonian tensor can be obtained both in real and reciprocal space. For simplicity, we have chosen the $\Gamma$ point to visualize the matrix shown in Fig.~\ref{fig:hopping} for two representative MOFs: \ce{Fe Pt C8 H4 N6} (or \texttt{qmof-3dfbcbd}) and \ce{Cd Ni C8 H12 N6} (or \texttt{qmof-4d9a98c}). The coefficients are always real numbers at the $\Gamma$ point, however, we might obtain complex numbers at other \textbf{k}-points. In the graphics, we plot the modulus-squared TB parameters $|t_{ij}|^2$. The localized atomic orbitals are the valence orbitals within the pseudopotential for every atom in the unit cell, which typically are \textit{s}, \textit{p} and \textit{d} orbitals. Thus, the TB matrix can be divided into blocks, where diagonal blocks represent interactions among the orbitals of the same chemical element, and off-diagonal blocks represent the interactions between different valence orbitals of different elements. 

The false-color image visualizes the strength of the TB parameters, which is an indicator of nearest-neighbors interactions. By looking at Fig.~\ref{fig:hopping}(a), the TB parameters correctly indicate the interaction, or hybridization, of Fe-N, N-C, C-H and Pt-C, as can be verified with the corresponding MOF structures. In Fig.~\ref{fig:hopping}(b), the interactions are Ni-C, C-H, C-N, Cd-N and N-H. Note, that we have computed both $U$ and $V$ parameters for the Ni atoms and the Ni-N bonds in the MOF structure shown, however, we have not performed any Hubbard calculations for the Cd atoms.

Since interactions among intra-site orbitals might be strong, the maximum value of $|t_{ij}|^2$ in the color bar has been set to 0.5 for facilitating the data visualization. The absolute maximum values are 121.8 eV$^2$ and 42.4 eV$^2$ for \texttt{qmof-3dfbcbd} (Fig.~\ref{fig:hopping}(a)) and \texttt{qmof-4d9a98c} (Fig.~\ref{fig:hopping}(b)), respectively. Based on the results obtained, we see that the representation of MOFs in TB lattice Hamiltonians can provide useful information on MOF topology and hybridization.

\begin{figure}[H] 
    \centering
    \includegraphics[width=1.\textwidth]{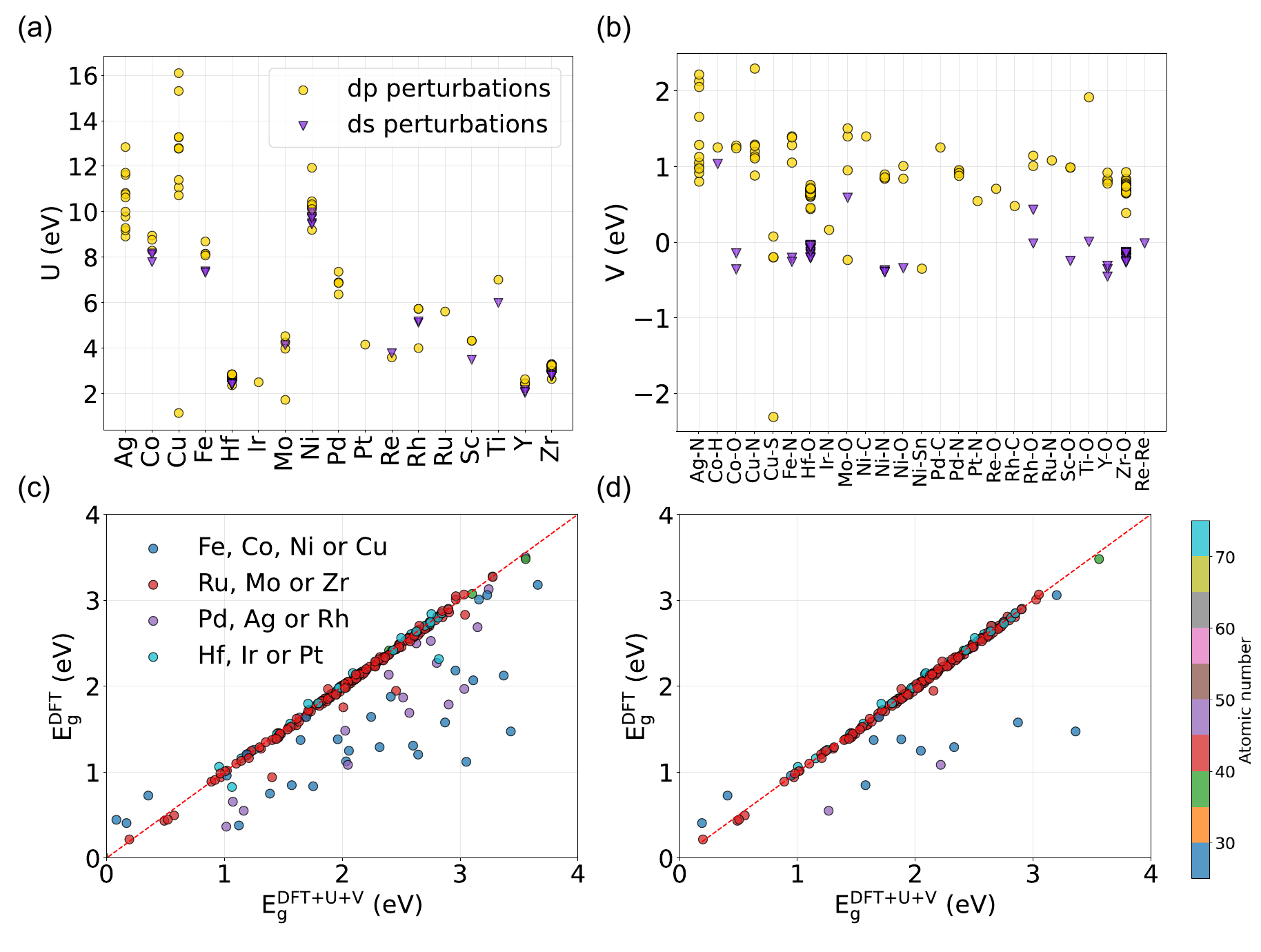} 
    \caption{Scatter plot of the intra-site $U$ and inter-site $V$ Hubbard parameters for the Extended Hubbard (EH) data set, ordered by the occurrence of transition metals in each material, considering the calculations with different type of manifolds, i.e., performing perturbations on (a) \textit{d-p} or (b) \textit{d-s}  orbitals, see Methods section for details.
    Scatter plot of band gap energies computed using standard DFT (E$_{\text{g}}^{\text{DFT}}$) and the DFT+U+V (E$_{\text{g}}^{\text{DFT+U+V}}$) framework for the EH subset considering (c) \textit{d-p} and (d) \textit{d-s} perturbations. The color map represents the atomic number of the transition metal associated with each material. }
    \label{fig:hubb_params}
\end{figure}

\textbf{Hubbard parameters.}
Modeling MOFs using the EH Hamiltonian requires not only the TB parameters, but also the intra-site $U$ and inter-site $V$ Hubbard parameters. $U$ is associated with a transition metal \textit{d}-orbital and $V$ refers to the interaction between the transition metal \textit{d}-orbital and one of its nearest-neighbor orbitals. We have provided two sets of values for $U$ and $V$, here defined as \textit{d-p} and \textit{d-s} perturbations referring to the manifolds perturbed. In Fig.~\ref{fig:hubb_params}(a), the distribution of $U$ values is plotted for each material of the EH data set. While a few structures may contain more than one transition metal species, we only plot one metal per MOF for simplicity. Depending on metal species, the intra-site parameter can have a large dispersion, as is the case for Ag and Cu. Or it can have very similar values in different environments, such as in the case of Zr, Hf and Y. While $U$ generally refers to the same orbital \textit{d}, performing the inter-site perturbation on \textit{p} or \textit{s} orbitals for computing $V$ can alter the $U$-outcome, yielding systematically smaller values for \textit{d-s} perturbations.

The $V$ distribution per metal-organic interaction is shown in Fig.~\ref{fig:hubb_params}(b). For clarity, we plot one $V$ value per structure, which is equivalent to the average nearest-neighbors $\langle V\rangle $. The data exhibits a large dispersion, where we observe positive values for \textit{d-p} perturbations, and relatively small, negative values for \textit{d-s} perturbations. 

The sets of $U$ and $V$ values provided in this work are aimed at supporting tight-binding modeling of metal-organic frameworks. They can be utilized in the training of machine-learning models and for exploratory data analysis. In addition, the EH data might support applications of quantum computing. In one example, the data are used as input for computing the band gap of representative semiconductors in a quantum-centric materials simulation workflow \cite{Duriez2025}. 

\textbf{Band gap predictions.}
Typically, the DFT+U+V methodology is applied to correct for the underestimation of band gap energies by standard DFT, which can fail to produce reliable predictions in systems with strong electronic correlations \cite{Pavarini2021}. This situation could occur in MOFs containing transition metals, localized orbitals and stronger electronic correlations and could be explored with our data contribution.

In Fig.~\ref{fig:hubb_params}(c)-(d), we show a comparison of the band gap energies computed using DFT and DFT+U+V for \textit{d-p} and \textit{d-s} perturbations, respectively. In both cases, we observe a large concentration of materials in the diagonal, indicating that band gap energies of most structures remain unaffected by DFT+U+V corrections. While this result is surprising, we note that most materials in the data set contain Zr or Hf. For other materials the band gap systematically increases, as expected. For additional information, we refer the reader to the Supporting Information.

\textbf{Applications in generative AI.}
The combination of TB embeddings and SOAP descriptors offers a powerful approach for generative materials discovery. In this framework, TB embeddings serve as a compact representation of the electronic structure, where a generative model would explore the parameter space proposing novel structure. SOAP descriptors could assist in reconstructing structural information, acting as an auxiliary tool for decoding atomic arrangements that may not be captured by the TB embeddings alone. Overall, this allows for predicting interactions and elemental species based on the TB embeddings  while simultaneously leveraging the SOAP descriptors for refining structural details.

For investigating this scenario, we have analyzed the metal atoms present in the TB data set. We have grouped atoms that are equivalent by symmetry, including those that appear distinct based on their CIF files if they are occupying nearly equivalent sites. From each group, we have selected one representative atom to ensure that the dataset remains balanced, avoiding over-representation of redundant entries.

\begin{figure}[t] 
    \centering
    \includegraphics[width=0.95\textwidth]{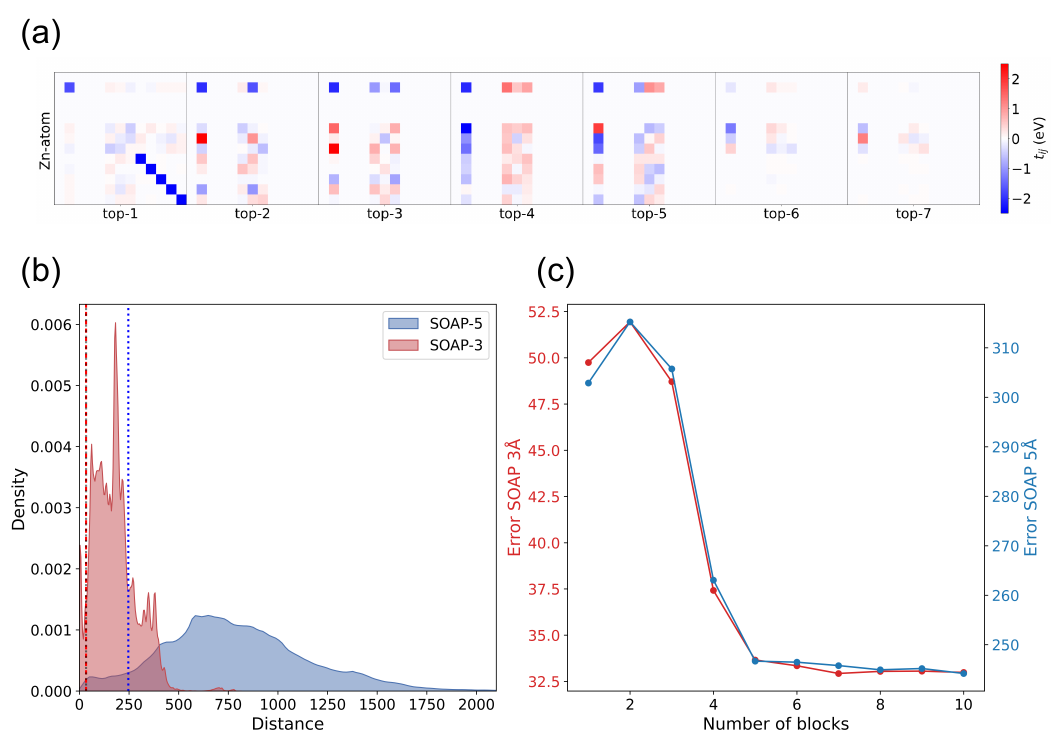} 
    \caption{
    (a) TB embeddings for a Zn atom in \texttt{qmof-fffeb7b}, showing the top-7 strongest interaction blocks extracted from the Hamiltonian parameters. Each block is represented by a $13 \times 13$ matrix and ranked by the absolute maximum value within the block, capturing the most significant electronic interactions regardless of sign. The $y$-label indicates the atom for which the embedding is computed, while the $x$-label denotes the rank in the $i\text{–}j$ block. To aid visualization, the values are plotted within the range $[-2.5, 2.5]$, with any values outside this range clipped by the color palette.
    (b) Distribution of pairwise Euclidean distances between SOAP feature vectors computed with SOAP-3\AA~ (red) and SOAP-5\AA~ (blue). Vertical dotted lines indicate the mean averaged error of the Euclidean distance in test set predictions, located at 33.346 for SOAP-3\AA~ (red) and 246.464 for SOAP-5\AA~ (blue).
    (c) Mean Euclidean Distance Error between true and predicted values (computed over all test samples) as a function of the number of Tight-binding embedding blocks used. The error is reported for both short-range (SOAP-3\AA~) and long-range (SOAP-5\AA~) descriptors.}
    \label{fig:Figure-5-combined}
\end{figure}

We have constructed the TB embeddings by selecting the blocks that contain the six strongest interactions for each metal atom, see Methods section for details. Each embedding consists of six $13 \times 13 $ blocks, resulting in a total vector size of 1014. The dataset contains 21,186 entries. In Fig.~\ref{fig:Figure-5-combined}(a), we show a visualization of the TB embedding for a specific atom. 

We have trained a \texttt{RandomForestRegressor} \cite{Breiman2001} for predicting SOAP descriptors based on TB embeddings in a four-dimensional PCA-reduced space. To that end,  we have used the reduced TB embeddings as input features and the full SOAP vectors as targets. For testing, the same PCA transformation, computed on the training set, is applied to the unseen test samples before making predictions. This approach ensures that no information from the test set is leaked into the training process. The same settings are applied to the prediction of SOAP-3\AA and of SOAP-5\AA descriptors. We have used an 85:15 train-test split and set the number of estimators to 100. Further details are provided in the Methods section.

In Fig.~\ref{fig:Figure-5-combined}(b), we show the distribution of pairwise Euclidean distances obtained for SOAP feature vectors using cutoff radii of 3 and 5 \AA~, respectively, in red and blue. The vertical dotted lines mark the mean error of the Euclidean distance in the test set predictions using the six strongest blocks. The error made for SOAP-3\AA~ falls within the lower range of its overall distance distribution, indicating that the predicted embeddings remain relatively close to their true values. For SOAP-5\AA, the error is slightly higher, as expected due to the larger environment being predicted. However, it still falls within the lower part of its distribution and provides an acceptable level of predictive accuracy.

To assess the predictive power of TB embeddings, we have progressively increased the number of included blocks from 1 to 10 and applied PCA to each variation. This approach allows us to evaluate how many blocks are necessary to achieve accurate SOAP predictions.   Fig.~\ref{fig:Figure-5-combined}(c) shows the mean euclidean distance error between the predicted and actual SOAP vectors as function of the number of included blocks. We observe that the error decreases significantly when increasing from 3 to 6 blocks. However, it remains fairly constant by further increasing the number of  blocks. This suggests that six blocks capture sufficient information for predicting SOAP descriptors. 

The TB embeddings can be used to identify species involved in interactions. When combined with SOAP descriptors, TB embeddings enable the resolution of the entire material composition. In the context of MOFs, the SOAP representation can be employed for searching similar structures within the existing MOF building blocks, i.e., metal clusters andorganic linkers. This would enable the reconstruction and validation of MOF structures through simulations, facilitating material property optimization as well as structural analysis. In the Supporting Information, we present an example illustrating this process in detail.

\textbf{Prediction of Hubbard $U$ and $V$ Parameters from TB Embeddings}

In the following, we assess the utility of tight-binding (TB) embeddings for predicting the Hubbard intra-site $U$ and inter-site $V$ parameters. In our setup, each $13 \times 13$ TB matrix block serves as input for a regression model predicting Hubbard parameter values. 

The intra-site Hubbard $U_{i}$ parameter corresponds to diagonal blocks representing electron interactions on atom $i$. The inter-site $V_{ij}$ parameters are derived from off-diagonal blocks representing interactions between atoms $i$ and $j$ within the same unit cell. For simplicity, we treat the intra-site Hubbard parameter $U_{i} \equiv V_{ii}$ as a special (diagonal) case of the inter-site parameter. For each MOF, we have selected the 10 strongest TB blocks -- ranked by the magnitude of their largest matrix element -- for building the training data set.

For avoiding ambiguity in periodic systems, we restrict our analysis to Hamiltonians evaluated at the $\Gamma$ point and apply the \textit{minimum image convention}. This means that for any given atom $i$, interactions are considered only if the associated atom $j$ lies within a closer distance than any periodic image of $i$. The cutoff ensures that only physically relevant, short-range interactions are included.

\begin{figure}[H] 
    \centering
    \includegraphics[width=0.96\textwidth]{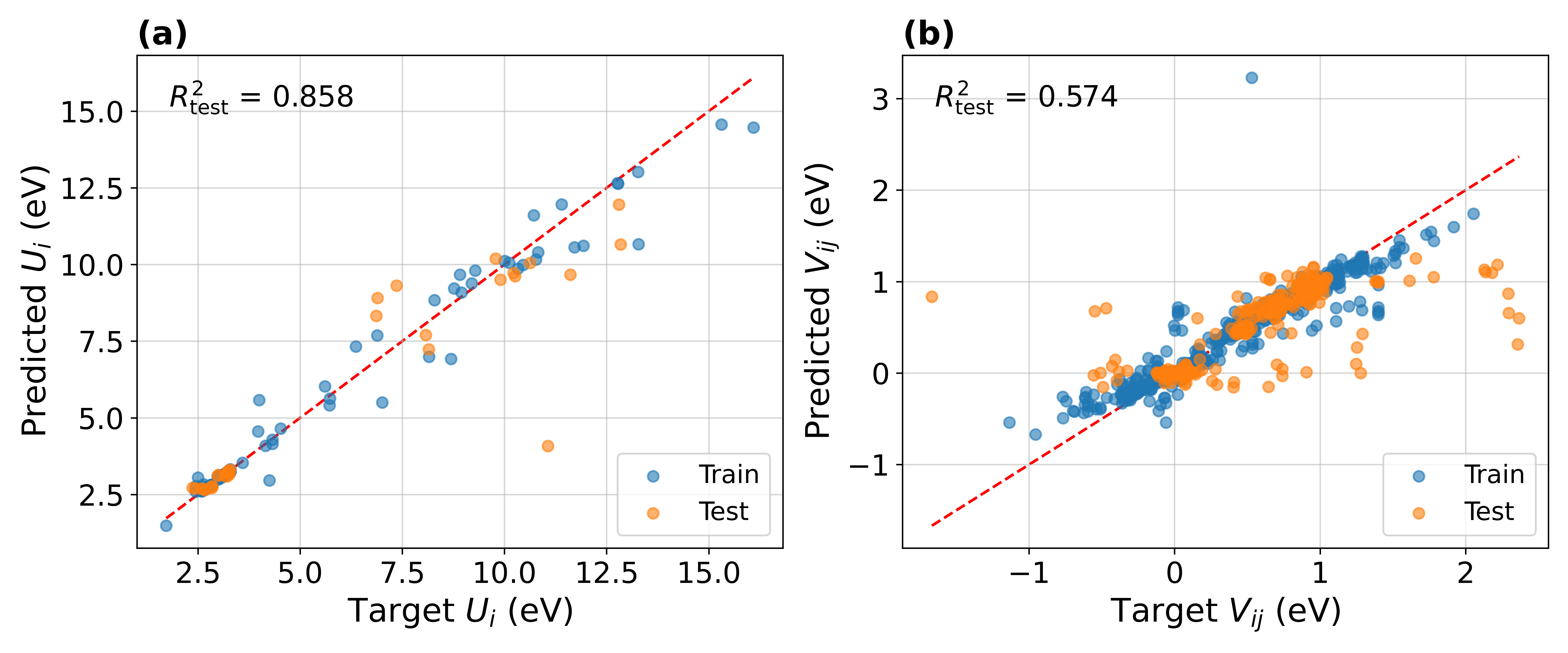} 
    \caption{
    Target and predicted Hubbard parameter values, for train (blue) and test (orange) sets, for one of the ten cross-validation splits.
    \textbf{(a)} Intra-site Hubbard parameters $U_i \equiv V_{ii}$.
    \textbf{(b)} Inter-site Hubbard parameters $V_{ij}$, for $i \ne j$.
    }
    \label{fig:prediction}
\end{figure}

The training data is taken from the \texttt{extended\_hubbard\_model/dp\_perturbations} subfolder. It contains 240 unique MOFs, lacking two MOFs with anomalously high $V$-values that we excluded. For each MOF in the data set, we have selected the $U$ and $V$ parameters corresponding to the first metal atom appearing in the structure. Of the original 9,754 entries, we have selected those 2,386 with the highest, absolute $V$ values per MOF. We have then performed the data splitting at MOF-level for ensuring that the test sets contained unseen materials, thus preventing data leakage.

We have trained a single \texttt{RandomForestRegressor} model for predicting both $U$ and $V$ values, without distinguishing between them. To validate the model's predictive capabilities, we have performed a 10-fold cross-validation using the top 10 strongest TB embedding blocks per MOF. We have trained the model with 100 estimators and default hyperparameters, leveraging the embeddings as input features and the Hubbard $V$-values as targets. We have implemented the cross-validation in Python using \texttt{scikit-learn}'s \texttt{RandomForestRegressor} and \texttt{KFold} utilities. Further implementation specifics, such as the selection of top-k blocks and data pre-processing, are described in the Methods section.

Even though data is limited, we observe that the model exhibits reasonable predictive performance. The average coefficient of determination $R^2$ is 0.914, with minimum and maximum values of 0.716 and 0.989, respectively. We obtain an average test mean absolute error (MAE) of 0.134 and mean squared error (MSE) of 0.179 across folds.

The modeling results are shown in \autoref{fig:prediction}. Overall, they demonstrate that the TB blocks carry the critical information with regards to electronic interactions in MOFs, and that both $U$ and $V$ values can be predicted robustly based on TB embeddings across a chemically diverse set of MOF structures.

\section*{Usage Notes}

The TBHubbard database contains 10,863 electronic structure simulations of MOFs that are split between the TB and EH subsets. In the TB data set, the PAOFLOW projection outputs are made available in a NumPy array for 10,435 MOFs as individual files. Each file contains the specific QE inputs and outputs, SOAP descriptors, and  Bader/QE electronic charge distributions. The SOAP descriptors, which capture the local atomic environments within MOFs, come in two variations: SOAP-3 \AA{} and SOAP-5 \AA. The descriptors represent the atomic topology at different length scales. The TB embeddings for each atomic species in each MOF are computed at the $\Gamma$ point. For visualizing the data and input generation, we have included a collection of auxiliary Python scripts. Specifically, we have provided scripts for generating self-consistent QE calculations which are based on CIF files and metadata within the QMOF database. In addition, we have made available the scripts for extracting the TB projection, for visualizing the projection matrix at $\Gamma$ points and for constructing the TB embeddings. The code contributions are designed to ensure reproducibility of the computational workflow. 

For consistency, the EH data set contains 428 files with the computation of the Hubbard parameters through construction of the susceptibility matrix, the corresponding set of $U$ and $V$ values, as well as the QE input/output for DFT+U+V calculations. For enabling accessibility, we have compiled JSON files containing the inputs and outputs of the QE calculations. In addition, Python scripts are included for reproducing the graphics, for generating JSON files as well as  the QE input files used in the ground-state calculations.

\section*{Code Availability}

Auxiliary scripts for creating the datasets, for generating the QE input files, and for plotting the figures are available in the database repository \href{https://dataverse.harvard.edu/dataverse/tbhubbard}{https://dataverse.harvard.edu/dataverse/tbhubbard} \cite{tbhubbard}. 

\section*{Acknowledgments}

We acknowledge funding (grant numbers 180544 and 225147) by NCCR Catalysis, a National Centre of Competence in Research funded by the Swiss National Science Foundation.  We thank Gavin Jones (IBM) for project support and Aleksandros Sobczyk (IBM) for the compilation and optimization of the Quantum Espresso code in several HPC architectures. Also, we would like to thank Ramon Cardias (CBPF) for introducing the PAOFLOW software to us and Marcio Costa (UFF) for the fruitful discussions and guidance on the use of PAOFLOW.

\section*{Author contributions}

P.C.C. developed the simulation workflow, created the dataset and wrote the manuscript. F.Z. developed the simulation workflow, TB embeddings, and predictive models, created the dataset and wrote the manuscript. A.C.D. and M.A.B. defined the database output requirements. R.N.B.F. developed the simulation workflow, analyzed the data and wrote the manuscript. B.J. (\textit{in memoriam}) and B.W. proposed the creation of the database. M.S. proposed the creation of the database, analyzed the data and wrote the manuscript.

\section*{Competing interests}

The authors declare no competing interests.

\bibliography{sn-bibliography}

\newpage

\section*{Supporting Information}

\renewcommand{\theequation}{S\arabic{equation}}
\renewcommand{\thefootnote}{\alph{footnote}}
\renewcommand{\thesection}{S\arabic{section}}
\renewcommand{\thefigure}{S\arabic{figure}}
\renewcommand{\thetable}{S\arabic{table}}
\setcounter{equation}{0}
\setcounter{section}{0}
\setcounter{figure}{0}
\setcounter{table}{0}

\begin{figure}[H] 
    \centering
    \includegraphics[width=1.\textwidth]{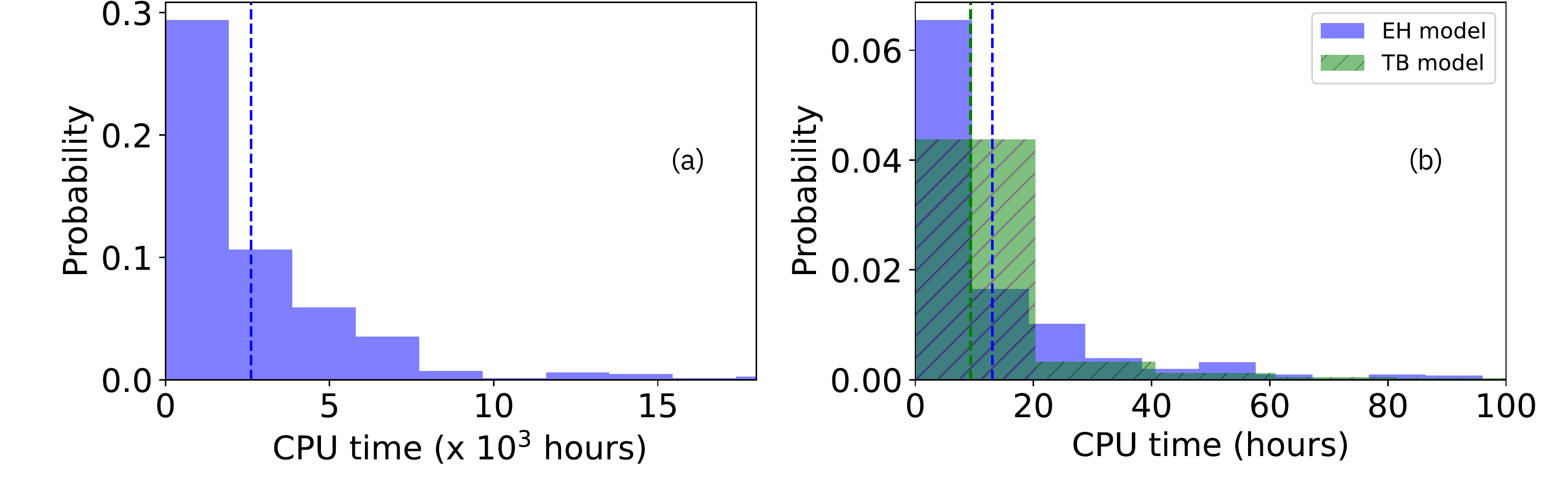} 
    \caption{Normalized histograms representing the CPU time in hours of the (a) Hubbard parameter computation using the executable \texttt{hp.x} from QE and (b) ground state calculation using the executable \texttt{pw.x} from Quantum Espresso (QE). Both subsets are accounted here, with 428 and 10345 calculations for the Extended Hubbard (EH), in blue, and Tight-binding (TB) subsets, in green hatched, respectively.  }
    \label{fig:cpu_time}
\end{figure}

The distribution of the CPU run time for the TB and EH subsets, considering the executing of \texttt{hp.x} and \texttt{pw.x}, is plotted Fig.~\ref{fig:cpu_time}(a) and Fig.~\ref{fig:cpu_time}(b), respectively. The average runtime is represented as a dashed vertical line. Note the different orders of magnitude for the executions, where \texttt{hp.x} is extremely more costly than the \texttt{pw.x}.

\begin{figure}[H] 
    \centering
    \includegraphics[width=1.\textwidth]{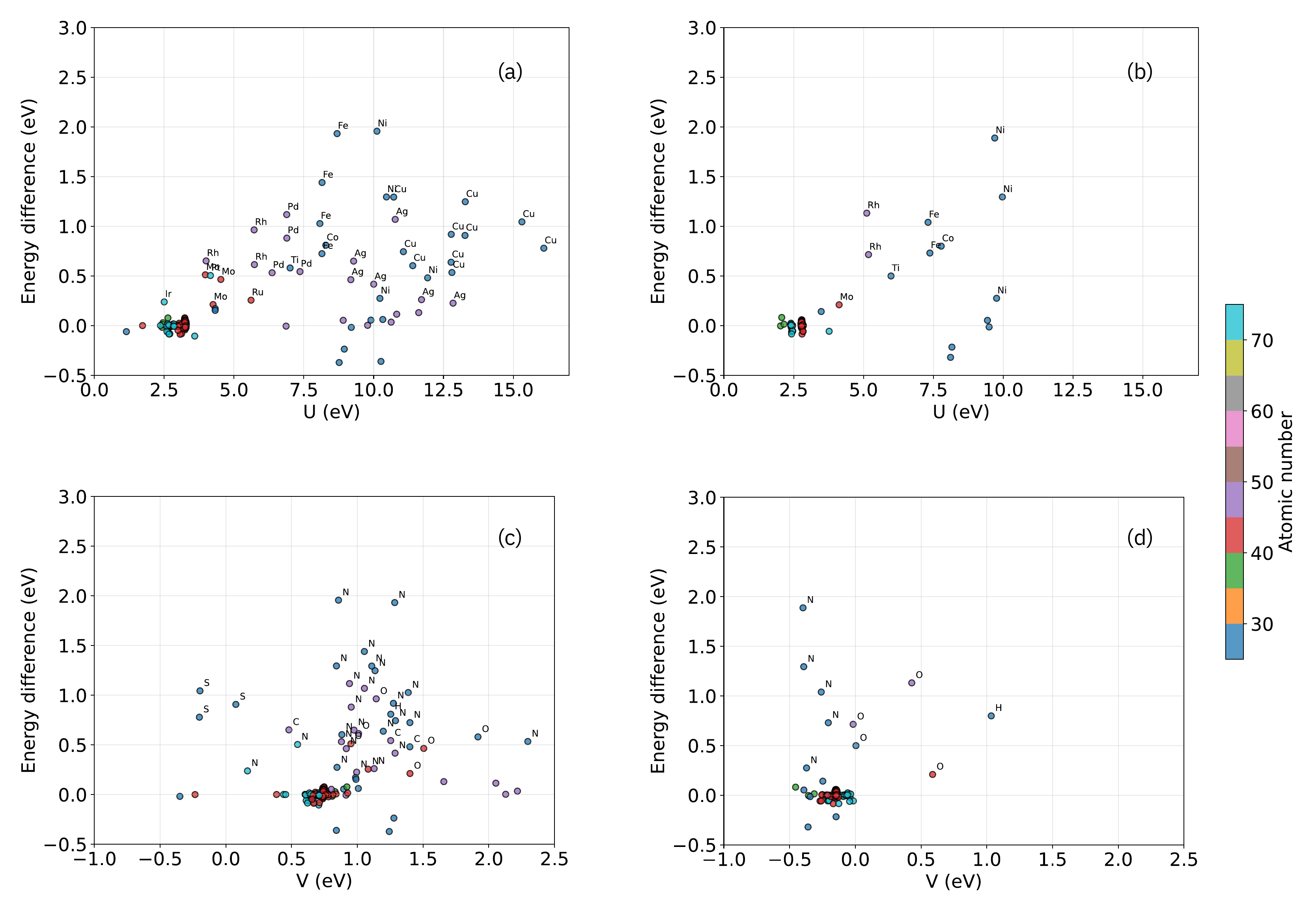} 
    \caption{Energy difference between band gaps computed with DFT (E$_{\text{g}}^{\text{DFT}}$) and DFT+U+V (E$_{\text{g}}^{\text{DFT+U+V}}$) as a function of U for dp (a) and ds perturbations (b) in the Extended Hubbard (EH) subset. The same is plotted as a function of V for dp (c) and ds perturbations (d), as well. The colormap represents the atomic number of the transition metal in each MOF.}
    \label{fig:bandgap_diff_UV}
\end{figure}

The energy difference between the DFT and DFT+U+V band gaps are plotted as function of $U$ and $V$ in Fig.~\ref{fig:bandgap_diff_UV}, for \textit{d-p} and \textit{d-s} perturbations. While a correlation might be drawn between the band gap energy difference and $U$, for $V$ this behavior is not as clear. 

\begin{figure}[H] 
    \centering
    \includegraphics[width=0.6\textwidth]{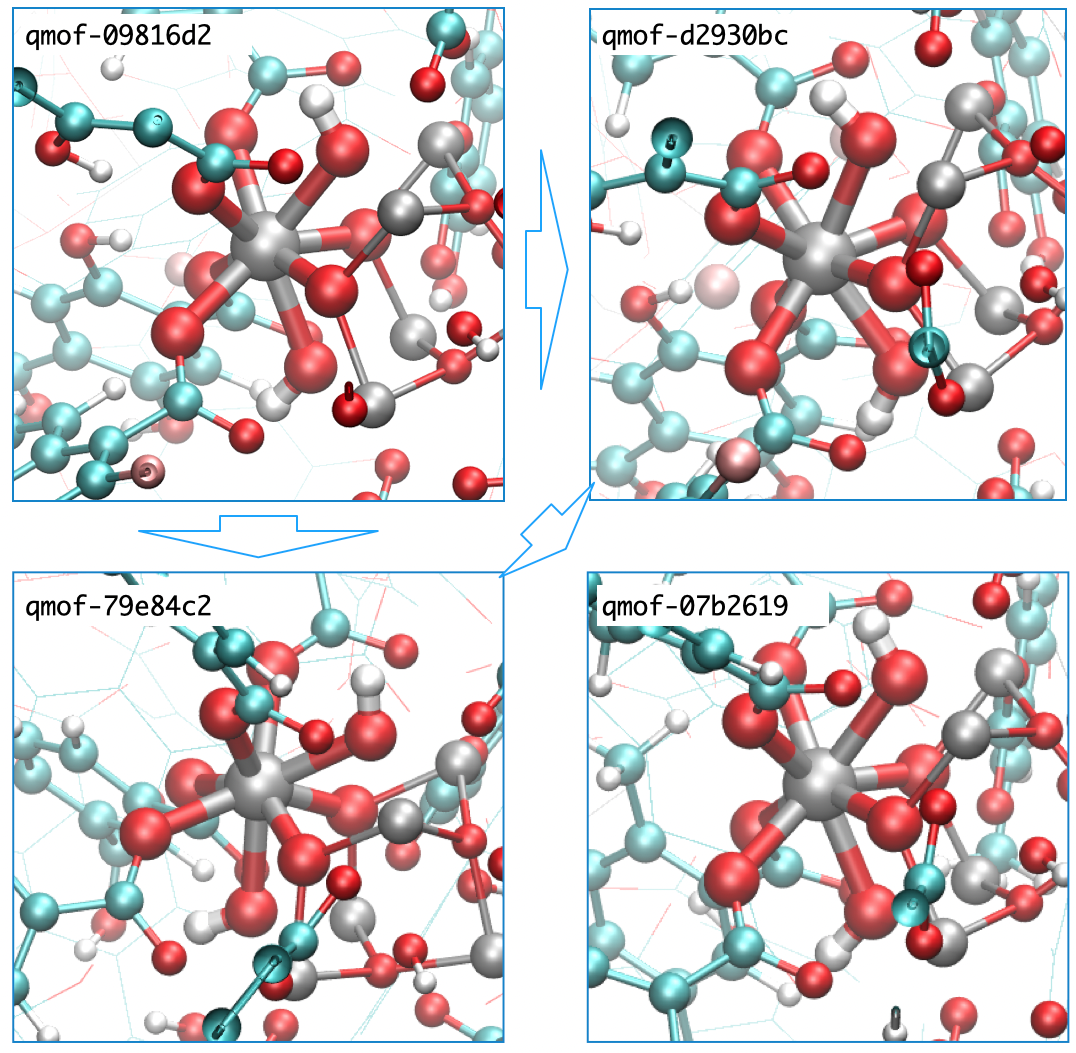} 
    \caption{
    Illustration of the initial structure from which the Tight-binding and SOAP embeddings search was conducted. The top left panel shows the selected metal atom within the MOF. Zr-atom in the \texttt{qmof-09816d2} (only present in the test set).
    The other three panels indicated by the arrow display the closest MOF structures based on the TB embeddings, highlighting the structural similarities to the initial configuration. Zr-atom in \texttt{qmof-d2930bc}, \texttt{qmof-79e84c2}, and \texttt{qmof-07b2619}.
    }
    \label{fig:TB_embeddings_similarity}
\end{figure}

{\bf Example of Structural Search Using TB and SOAP Embeddings.}
To demonstrate the utility of TB and SOAP embeddings, we begin by focusing on the TB embeddings of metal atoms across a specific MOF, as previously restricted for example purposes in the context of metal cluster design.
Using the same train-test split from Fig.~\ref{fig:Figure-5-combined}, from the main text, we select one metal from the test set, for which we have the ground truth. For this test, we treat the corresponding TB embedding vector as if it were predicted by a genetic algorithm. We then search for the closest TB embeddings within the training set and visually compare the resulting structures, as shown in Fig.~\ref{fig:TB_embeddings_similarity}.
This search reveals a significant structural resemblance between the predicted and true structures.

\begin{figure}[H] 
    \centering
    \includegraphics[width=0.6\textwidth]{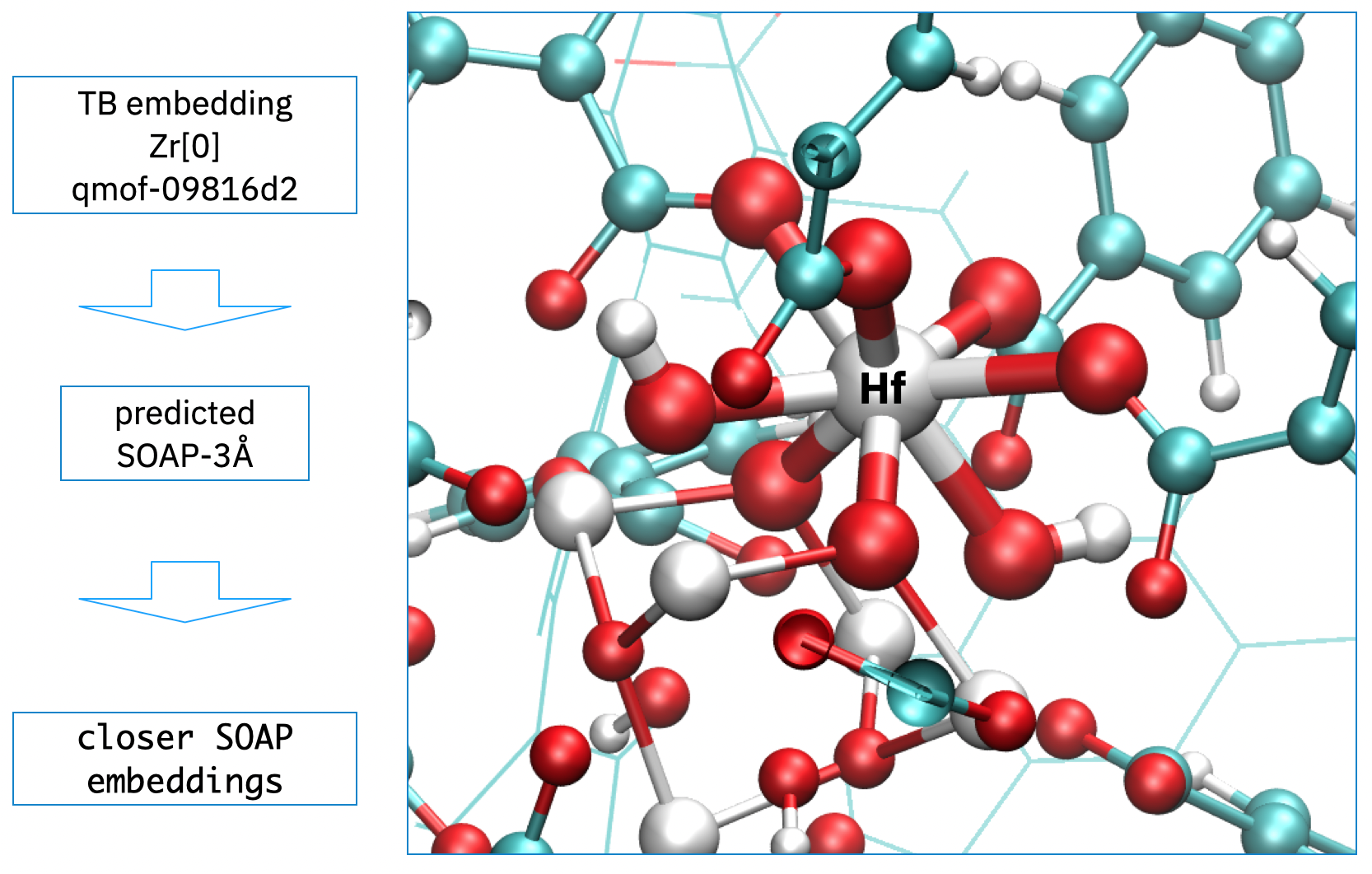} 
    \caption{
    A \texttt{RandomForestRegressor} predicts the SOAP-3\AA~ embedding from the Tight-binding embeddings. 
    The right panel presents one of the closest structures identified in the SOAP space with a different metal atom, demonstrating the metal-agnostic nature of SOAP descriptors. In this case, we identify an Hf atom in \texttt{qmof-75cdf73} as having a highly similar local environment.
    }
    \label{fig:SOAP_embeddings_similarity}
\end{figure}

Simultaneously, we predict the SOAP descriptors for this same metal atom. Since the corresponding SOAP descriptor is available in the ground truth of the test set, we compare the predicted SOAP descriptor to the true one, which differs only by a small error for SOAP-3\AA~ and a marginally larger error for SOAP-5\AA.
We proceed to search for the closest vector in the SOAP space, which is nearer to the predicted descriptor rather than the true one. By doing so, we can identify similar Zr atoms across different MOFs, see Fig.~\ref{fig:TB_embeddings_similarity}.
Furthermore, since SOAP descriptors are built with a reduced species set that maps all metals to a single category, we can perform a metal-agnostic query. This allows us to search for MOFs that exhibit the same local environment but with a different metal species, such as Hf, which we identify in the results, see Fig.~\ref{fig:SOAP_embeddings_similarity}.

\end{document}